# Switching dynamics of morphology–structure in chemically deposited carbon films –a new insight

**Mubarak Ali, [a],* and Mustafa Ürgen[b]**

[a] Department of Physics, COMSATS Institute of Information Technology, Islamabad 45550, Pakistan.

[b] Department of Metallurgical and Materials Engineering, Istanbul Technical University, 34469 Maslak, Istanbul, Turkey.

*corresponding address: mubarak74@comsats.edu.pk, mubarak74@mail.com , Ph. +92-51-90495406

**ABSTRACT** –Carbon is one of the most investigated materials and shows chaotic behavior in terms of evolving structure. Synthesizing carbon materials largely depend on the deposition technique, process parameters, condition of substrate surface and ratio of the gaseous chemistry. A variety of techniques have been employed to depositing carbon films from various gaseous mixtures to different substrate materials. In this study, carbon thin and thick films are discussed for different techniques known as hot filament chemical vapor deposition and microwave plasma chemical vapor deposition where their synthesis process has been explained in a new context. Here, we discuss attained dynamics of atoms (or their tiny grains) amalgamating into a particular phase of grain or crystallite and electron-dynamics responsible for binding atoms in the formation of all sorts of tiny grains, grains and crystallites controlling overall morphology-structure of films thickness at few nanometers to several microns. Carbon atoms when in solid state, on amalgamation at flat surface result into bind under uniform electron-dynamics and when the amalgamation is at uneven surface, (even at atomic level) they result into bind under non-uniform electron-dynamics. Where binding of atoms is at uniform electron-dynamics, a graphitic structure evolves following by different modifications into other carbon phases depending on the orientation of electron states with respect to centre of inner part of atom known as nucleus. Substrates under appropriate surface defects or abrasion result into an improved rate of nucleation of tiny grains, hence, their increased rate of growth. This study embarks on unexplored science of carbon films where in addition to localized process parameters nature of substrate also influence dynamics of formation of tiny clusters, grains and crystallites at their initial stage of formation. Our results and discussions enlighten us to revisit the nucleation and growth mechanisms of different sorts of films deposit at any scale and at any substrate surface constituting different composition.

***Keywords:*** Carbon films; Process parameters; Dynamics; Deposition; Morphology-structure; Grains; Crystallites; Field force

**1. INTRODUCTION**
Recognition of an element is based on atomic number and in the Periodic Table it provides the position depending on the nature of electronic states. Elements where atoms possess filled states should work in a different way to ones with unfilled state (s). The occupancy of only one unfilled state in the case of hydrogen atom leads into different behavior compared to those comprised two broad categories; atoms with filled states and atoms with unfilled state (s). In carbon atom, not all of the outer eight states are filled and having the deficiency of four electrons according to Periodic Table. This attribute of carbon atom allows us to treat it gas as well as solid depending on the orientation of sets of electrons around the inner part of the atom known as nucleus [1]. In this context, the key factor to evolve carbon structure (master) should be the certain orientation of sets of electrons around the nucleus following by their varying placement angles around the nucleus and with respect to centre of nucleus made under certain arrangement of the process, thus, resulting into emerge modified phase of master structure, one at a time depending on the mechanism of certain arrangement.

It has been disclosed that gold atoms amalgamate under attained dynamics following by their binding under electron-dynamics [2]. In such atoms where electrons, on excitation de-excite to restore the positions (states), photons shape-like Gaussian distribution are resulted and is the cause of binding atoms in any sort of structure [3]. A tiny shaped particle has been discussed along with elongation and modification into smooth elements [4]. A carbon film, where tiny grains possess two-dimensional structure (graphite structure or master structure) has been discussed along with enhanced field emission characteristics on elongation and modification of tiny grains into smooth elements, known as UNCD/NCD film [5]. The formation mechanism of different tiny particle under varying concentration of gold precursor has been discussed elsewhere [6]. Again, efforts have been made to tap tiny shaped particles of silver, binary composition of gold and silver and gold in pulse-based electronphoton-solution interface process while processing their precursors where it was concluded that nature of precursor is the one, directing conditions for the formation of large size shaped particles [7]. A tiny particle changing shape and size under varying ratio of pulse OFF to ON time has been investigated elsewhere [8]. A diffusion mechanism of atoms and tiny-sized particles starting from the stretching of electron states of atoms has been discussed while developing extended shape particles under unprecedented fast rate of their development [9] where the law of reflection also included the orientation of surface made into smooth elements dealing incident light and reflected light. Atoms of suitable elements belonging to metals and semi-metals group experience electron transitions and they do elongate or deform but not ionize while inert gas atoms split under suitable field of photonic current [10] and revealing the phenomenon of heat energy and photon energy has been discussed elsewhere [11], in which silicon atom was considered to be the intermediate component for regulating the energy. A tiny-sized particle capable to work either as an effective nanomedicine or as a defective nanomedicine has been discussed elsewhere [12].

In most cases, to deposit and synthesize material, it is extracted from the precursor. Different processing approaches have been utilized for the synthesis of various carbon materials under the variation of process conditions, on examining, revealed different characteristics regardless of that atomic nature of the targeted material (carbon) can never be changed in an isolated system due to naturally built-in machine as per fundamental laws; both energy and mass is remained conserved. Again, special emphasis was remained on the morphology-structure change in those materials under varying process parameters. Despite of that, what nature of the source precursor an atom dealt in the course of dissociation or detachment is to be autonomous, the affinity with counterpart either to atoms of the similar nature or different nature is crucial while targeting needed material evolving certain phase of structure; it also invokes fundamental question, "how atoms of carbon evolve structure and switch electron-dynamics resulting into modification of the phase into those known in their exceptional hardness"? It is necessary to address such fundamental questions of materials science while synthesizing carbon-based materials at nanoscale to microns as they set foundation of advanced engineering and then owing to varying properties (characteristics) under the length scale.

A large number of publications have been appeared on the fabrication of microcrystalline diamond films as well as nanocrystalline diamond (NCD) and ultrananocrystalline diamond (UNCD) films under varying conditions and their synthesis mainly involve techniques known as hot filament chemical vapor deposition (HFCVD) and microwave plasma chemical vapor deposition (MPCVD). A vast range of morphological features of such films observed under different process parameters have been published. However, due to greater availability of parameters range and enhanced growth rate, diamond films synthesized *via* HFCVD show more versatility in terms of morphological features. The morphology of the grains/crystallites in various diamond films targets some selective applications. For example, diamond films evolved with large crystallites (very large size grains) are strong candidates for heat sink application and free-standing diamond films for X-ray windows [13, 14], a film in average (reasonable) size of grain is considered suitable for application like high frequency loudspeaker diaphragms [13], a film in small grain size is suitable for cutting tools applications [13, 15] and a film in ultra-small size of grain is considered to be a strong candidate for field emission or display panel applications [16-18].

Physics and chemistry in processing carbon-based materials under a range of schemes have been explained in quite a number of studies. In HFCVD, growth mechanism of cubo-octahedral diamond is the competing growths of (100) and (110) crystallographic planes [19]. Model study of diamond films only validates the critical role of aromatic condensation and interconversion of carbon phases mediated by atomic hydrogen in gas-activated deposition [20]. In hot-filament reactor, transport of atomic hydrogen to the growing surface is diffusion limited process under commonly employed conditions [21]. Understanding CVD diamond growth is a complex phenomenon which makes modeling of diamond crystallization a challenging task [22]. In HFCVD and under specific conditions of argon gas environment, the transition of microcrystalline diamond film to nanocrystalline diamond film is observed [23]. On reducing the secondary nucleation, diamond coatings show a very high purity in Raman signal, thus, varying the gas pressure that expands the window of depositing films at high/low growth rate [24]. An improved model of growth mechanisms of diamond films grown *via* HFCVD both in $Ar/H_2/CH_4$ and traditional $CH_4/H_2$ gas mixtures give some useful information regarding carbon atoms and methyl radicals [25]. May and Mankelevich [26] developed a model for diamond crystallite sizes ranging from 10 nm to several millimeters where growth of diamond is a sliding scale between atomic hydrogen and hydrocarbon radical, and different growth conditions only serve to fix the resulting film morphology and growth rate. In HFCVD, growth rate of the diamond films is influenced jointly by substrate temperature and total pressure [27]. Effects of methane concentration on the characteristics of diamond films have been studied in both HFCVD [28] and MPCVD [29].

Mainly, UNCD/NCD films meant specifically for field emission applications were developed under varying process parameters, gas concentrations/dopants and also with composite/hybrid structures [16-18, 30-37]. In UNCD/NCD films, the encapsulated basic idea provides effective route to conduct heat while charge is conducted (flow of electrons) through transpolyacetylene layer around the boundaries of tiny grains. However, our recent studies reveal that in the course of attempting scientific goals considering conduction of charge (flow of electrons) is not a viable solution [5, 10] and conduction of heat energy along with propagation of photons characteristics current have been discussed elsewhere [11]. Again, a field force influences (forces) atoms from a distance depending on the nature of their electron states [1].

Our previous studies (also referred the work of others in this study) discussed the various aspect of processing thin thick films of diamond (mainly) along with graphite under various process conditions in the course of employing different substrates' materials and seeding treatment of substrate in chemical vapor deposition as referred in the text. However, the heart of underlying science of carbon-based materials remained crucial and peculiar since the birth of carbon; how, an evolving structure switches one phase to another one, in some cases restoring the phase back, and what are the implications of that phase with respect to neighboring ones, what are the binding mechanism of carbon atoms, their primitive cell evolution and last but not least features of thin or thick film deposited on different nature substrate. These are the questions, which intrigued all through the career of a materials scientist, and more than that to those worked/working on carbon-based materials. This study is an attempt to address many of them, in some cases with more detail, and in other cases some detail.

This study discusses the formation of tiny grains, grains and crystallites of thin/thick films both in terms of their surface feature as well as cross-sectional view, thus, pinpointing the trustworthy switching dynamics of their formation. Substrates' relation to depositing material was mainly expressed in terms of atoms' amalgamation and formation of tiny clusters (tiny grains). Although, this study move forward from bottom to top; atoms' amalgamation into primitive cell to tiny cluster called tiny grains (known as NCD or UNCD), tiny grains to grains and grains to crystallites (ballas) but may deal uneven flow in presentation due to inter-connecting and together debatable science of the materials. Overall, this study incite the same morphology-structure of carbon-based materials, which is being dealt since decades, presenting general point of view, and somehow, specific insight on the underneath science where carbon films synthesized on different substrates, treated differently, investigating morphology-structure of tiny grains, grains and crystallites in terms of attained dynamics of carbon atoms and electron-dynamics.

In this study, a variety of carbon thin and thick films have been synthesized at differently treated substrates' surface where morphology-structure investigated in terms of attained dynamics of atoms and electron-dynamics along with structural modifications. The fundamental process of formation of tiny grains, grains, and crystallites under varying process parameters along with switching phase of evolving structure under confined electron-dynamics has been discussed. Clearly, this study explores the unexplored science of the synthesis of carbon-based materials in the form of thin thick films and their explanations are based on different lines to the existing ones. This study pinpoints that amalgamation of atoms or their tiny grains into a particular structure is due to attained dynamics where binding of atoms (or sticking of tiny grains) at substrate surface along with their modification into different phases is due to electron-dynamics.

## 2. EXPERIMENTAL DETAILS

Microcrystalline/polycrystalline carbon films were synthesized on differently treated silicon and molybdenum-coated titanium substrates by employing deposition technique known as hot filament chemical vapor deposition (HFCVD). The explanation of underlying science of carbon films known in ultrananocrystalline diamond (UNCD) and nanocrystalline diamond (NCD) are mainly synthesized in technique known as microwave plasma chemical vapor deposition (MPCVD). However, the present study contained the experimental results is only related to HFCVD.

In the deposition of some of the carbon films, silicon wafers (p-type) of thicknesses around 400 µm supplied by Wafer World (USA) along with differently employed surface treatment were used as substrate material. Mo-coated titanium substrate was prepared by depositing molybdenum on titanium pieces in high power cathodic arc physical vapor deposition technique by which several articles and theses were published by the Professor Ürgen group. Substrate were prepared as per procedure given in SAMPLE A to G and placed over two centimeters thick molybdenum plate followed by fused alumina plates barrier, which were placed over the copper holder through which cool water was circulating, thus, maintained the thermal conductivity of the substrates. Volume of the chamber was approximately 2.9 cubic ft. After evacuation of the chamber upto ~ $10^{-3}$ Passel, different mass flow rate of gases were introduced to the reactor by controlling with mass flow controllers. The tantalum filaments were used as the hot filaments diameter 0.5 mm of each wire. The heating of substrate's stage was maintained through heat of hot filament.

The parameters studied while synthesizing a variety of carbon thin thick films are given below with SAMPLE name and so in the Figure captions of surface topography and fracture cross-section images. Further details of materials' specification, seeding treatment and the process are given elsewhere [14, 28, 38-41]. The preparation procedure of unseeded silicon substrates is given elsewhere [14], whereas, a detailed process of preparing molybdenum-coated titanium substrates and temperature measurement are given in our previous study [39]. Surface morphology and fracture cross-section images of various tiny-sized grains, average-sized grains and micron-sized crystallites of carbon deposited in the form of films were obtained by field emission scanning microscope known as FE-SEM (JEOL, Model: JSM-7000F). Films of different carbon-based materials were analyzed by micro-Raman spectroscopy (HR800 UV; 632.8 nm, He-Ne Red Laser).

**SAMPLE A**; total mass flow rate: 200 sccm (2% $CH_4$), chamber pressure: ~ 55 torr, growth time: 8 hours (nucleation time: 30 minutes where concentration of $CH_4$ was kept 3% and pressure was increased from 1 torr to 55 torr), distance b/w substrates and hot filaments: ~ 8.0 mm, substrate temperature: 650 ± 20°C, input power: ~ 3 (kW) and ultrasonic agitation of substrate for 30 minutes in acetone having 500-600 mesh diamond powder (80 ml:8 gram) following by 10 minutes wash with acetone in ultrasonic bath.

**SAMPLE B**; total mass flow rate: 200 sccm (1.5% $CH_4$), chamber pressure: ~ 40 torr, growth time: 9 hours & 30 minutes (nucleation time: 30 minutes where concentration of $CH_4$ was kept 2% and pressure increased from 1 torr to 40 torr), distance b/w substrates and hot filaments: ~ 8.0 mm, substrate temperature: 800°C ± 20°C (decreased from ~ 890°C), input power: ~ 3 (kW) and ultrasonic agitation for 90 minutes with 500-600 mesh diamond powder in acetone solution (7 gram:70 ml) following by 10 minutes wash with acetone in ultrasonic bath.

**SAMPLE C**; total mass flow rate: 300 sccm (2.2 % $CH_4$), chamber pressure: ~ 5 torr, growth time: 10 hours, distance b/w substrates and hot filaments: ~ 8.0 mm, substrate temperature: (first increasing from ~ 650°C to ~ 770°C then decreasing from ~ 770°C to ~ 670°C), input power: ~ 3 (kW) and substrate was mechanically scratched with 500-600 mesh diamond powder and 30-40 mesh diamond powder following by 10 minutes ultrasonic agitation in 5 μm sized diamond powder in acetone and washing in acetone for 10 minutes in ultrasonic bath.

**SAMPLE D**; total mass flow rate: 300 sccm (2.2 % $CH_4$), chamber pressure: ~ 65 torr, growth time: 10 hours, distance b/w substrates and hot filaments: ~ 8.0 mm, substrate temperature: (first increasing from

~ 680°C to ~ 780°C then decreasing from ~ 800°C to ~ 700°C), input power: ~ 3 (kW) and substrate was mechanically scratched with 500-600 mesh diamond powder and 30-40 mesh diamond powder following by 10 minutes ultrasonic agitation in 5 μm sized diamond powder in acetone solution and 10 minutes wash with acetone in ultrasonic bath.

**SAMPLE E**; total mass flow rate: 300 sccm (1.5 % $CH_4$), chamber pressure: ~ 55 torr, growth time: 10 hours, distance b/w substrates and hot filaments: ~ 8.0 mm, substrate temperature: 800°C ± 20°C, input power: ~ 4 (kW) and substrate was mechanically scratched with 500-600 mesh diamond powder and 30-40 mesh diamond powder following by 10 minutes ultrasonic agitation in 5 μm sized diamond powder in acetone solution and 10 minutes wash with acetone in ultrasonic bath.

**SAMPLE F**; total mass flow rate: 200 sccm (2% $CH_4$), chamber pressure: ~ 40 torr, growth time: 10 hours, distance b/w substrates and hot filaments: ~ 7.0 mm, substrate temperature: 900 ± 20°C, input power: ~ 3 (kW), increased temperature under the same input power achieved by placing the substrates at three different locations underneath the hot filaments, and ultrasonic agitation for 90 minutes with 500-600 mesh diamond powder in acetone solution (7 gram:70 ml) and 10 minutes wash with acetone in ultrasonic bath.

**SAMPLE G**; total mass flow rate: 400 sccm (3 % $CH_4$), chamber pressure: ~ 75 torr, growth time: 9 hours, distance b/w substrates and hot filaments: ~ 7.0 mm, substrate temperature: 900 ± 20°C, input power: ~ 5.5 (kW) and ultrasonic agitation with 5 microns diamond powder in acetone for 10 minutes (7 gram:70 ml) and 10 minutes wash with acetone in ultrasonic bath; scanning microscope surface image of large crystallites deposited on (c) unseeded Si substrate and (d) seeded Si substrate; total mass flow rate: 400 sccm (3 % $CH_4$), chamber pressure: ~ 80 torr, growth time: 9 hours, distance b/w substrates and hot filaments: ~ 7.0 mm, substrate temperature: 830 ± 20°C, input power: ~ 4.7 (kW) and ultrasonic cleaning with 5 microns diamond powder in acetone for 10 minutes (7 gram:70 ml) and 10 minutes wash with acetone in ultrasonic bath.

## 3. RESULTS AND DISCUSSION

In Figure 1 (a), scanning microscope surface image shows non-uniform distribution of carbon tiny clusters (tiny grains). The size of the tiny cluster is varied from ~ 10 nm (encircled by small circle) to ~ 50 nm (encircled by large circle). A similar trend of tiny cluster is observed in Figure 1 (b); however, some of the tiny clusters increased the size on consolidate, thus, altered morphology by increasing the size where the average size of the grain is in the range of 200 nm (encircled by rectangular-boxes). In Figure 1 (b), regions encircled under small circles reveal the adherence of tiny clusters growing into the bigger ones. Uneven distribution of tiny clusters on seeded Mo-coated Ti substrate is more related to rough surface and those tiny clusters (in Figure 1) where only few atoms attached (size < 10 nm) are difficult to identify under the application of scanning microscope. Tiny clusters increased their size while coordinating approached ones, thus, switched morphology into cauliflower shape (Figure S1); several tiny clusters amalgamated into isolated particles size ~ 0.5 µm and some are in very large size.

In Figure 2 (a), the upper most surface of grains reveals switching cauliflower-like morphology into cube shape morphology. However, due to different rate of lateral deposition of carbon atoms as compared to in-plane deposition, grains don't grow in complete cube morphology and their top surfaces don't reveal square shape as well. In Figure 2 (A), the scanning microscope fracture cross-section of film demonstrates that growth of grains started from tiny clusters. The tiny clusters developed under favorable dynamics of amalgamating atoms followed by sticking at favorable sites of substrate's surface where it gradually evolved into grains and crystallites having their faceted and smooth morphology as long as all the tiny grains had been grown; some of the grains/crystallites (size > 1 µm) reveal cube shaped morphology started evolving since nucleation indicating atomically flat occupied surfaces along with depositing at uniform rate. In fact, employing different process parameters resulting into change the morphology of grains and crystallites. As in the case of Figure 2 (b), due to secondary nucleation morphology of grains and crystallites is more like 'snow covering the rough surface' and similar trend in morphology of grains and crystallites is observed in fracture cross-section image shown in Figure 2 (B); an average size of crystallite is ~ 900 nm where carbon atoms appear to be deposited on initiation of secondary nucleation and such growth behaviors prevailed where localized temperature decreased from higher limit to lower limit (higher limit was ~ 890 C° and lower limit is ~ 820 C°). Now when the temperature decreased from higher limit to lower limit, the pressure of the chamber ascended resulting into slow down the growth process, thus, slow pace switching of the process of growing grains and crystallites toward the nucleation side again in that film. Some other related changes in the on-going parameters may be considered. Thus, that nucleation now becomes the secondary nucleation as it started at second stage of the process or it can be termed reverse nucleation as it gives the opposite sense to primary nucleation. In secondary nucleation, atoms re-start to deposit both in terms of their attained dynamics as well as electron-dynamics and leading toward controlled symmetry process so that a particular phase of structure could sustain for longer time under the process conditions close enough to the ones at the nucleation stage. The introduction of secondary nucleation process is quite common in HFCVD as it is difficult to maintain temperature of the process for several hours and increasing the temperature by using the available means of option, we are varying the process conditions at different time within the same experiment. Thus, to explain results under the variation of single parameter is crucial. This reveals an effective change in the evolving structure under the variation of temperature ($\Delta T \sim 100°C$). In Figure 2 (c), morphology of crystallites is in hail-like shape where the average size of the crystallite is $\sim 1$ µm, some of the tiny grains coalesced under their different attained (emerged) dynamics and morphology is more like cauliflower shape, some of them coalesced under symmetric attained dynamics and made bigger size where morphology of the crystallite is more like ballas (sphere). Similar morphological features of grains/crystallites are shown somewhere else [42, 43]. Pyramidal-like morphology of diamond crystallites is shown in Figure 2 (d). Initially, tiny clusters covered the abraded surface of silicon followed by uniform growth. The process parameters manipulated the attained dynamics of atoms and/or tiny grains in such a manner that scale of growing morphology of grains (crystallites) is more like pyramidal shape. Those grains that nucleated afterwards (in the left vacant regions of the substrate) grew in different morphology as compared to the ones in pyramidal shape where growth dynamics followed the different procedure. The size of bigger crystallites is larger than 3 µm where faces grew faceted as shown in Figure 2 (d). Growth kinetics of different planes of homoepitaxial diamond films have been discussed [44].

The morphology of grains and crystallites of carbon films in Figures S2 (a) and S2 (b) is the same but due to greater thickness of seeded Mo-coated Ti substrate (1 mm) they grew bigger as shown in Figure S2 (a) where substrate was closer to filaments, thus, developed under different attained dynamics of depositing atoms and/or tiny grains. However, due to smooth surface of seeded silicon substrate, the distribution of grains and crystallites is more uniform. Again, in Figure S2 (c) due to rough surface of seeded molybdenum-coated titanium substrate, only few grains and crystallites of the film reveal pyramidal-like morphology and more are in the broken edges. In Figure S2 (d), the grains and crystallites are in pyramidal-shaped morphology and a very large crystallite protruded from the film under different conditions of amalgamating atoms and tiny grains. Under the same conditions, a film with uniformly distributed pyramidal-shaped grains and crystallites is shown in Figure S2 (e) because the substrate surface dealt suitable and uniform level of seeding treatment by retaining the surface flat free from any sort of scratches and fracture and was not the case in Mo-coated Ti substrate. We can see the uniformity and flatness of seeded silicon surface in the fractured cross-section image of the scanning microscope fracture (in Figure S2E) where the deposition at interface is uniform and growth of film is in columnar shape (thickness ~ 650 nm). The transformation of grains' morphology from simple cube shape (in Figure S2b) to pyramidal shape (in Figure S2e) is mainly related to difference of nucleation rate of tiny grains under different conditions of temperature, concentration of methane and chamber pressure as printed in their Figures' caption. At different process parameters, dynamics of formation of tiny grains altered along with adherence to substrate resulting into transform the morphology of grains from simple cube shape to pyramidal shape. The similarly featured morphology of films is shown elsewhere [45] as in the case of Figure S2 (a), Figure S2 (b) and Figure S2 (e). Again, similarly featured morphology of films is shown elsewhere [46] as in the case of Figure S2 (a) and Figure S2 (b). In Figure S3 (a) and Figure S3 (b), the size of particles is ~ 400 nm, which is termed as crystallites where surface of the square face is smooth, however, surface of triangle-like face is not smooth. In Figure S3, the formation of isolated crystallites on the surface of highly abraded silicon substrate is related to high input power along with utilization of high ratio of hydrogen to methane concentration where uniform activation of electron-dynamics resulted matching to attained dynamics of prior amalgamated atoms at each side of square faces with unbroken symmetry.

In Figure 3 (a) and Figure 3 (b), coarse morphology of ballas crystallites evolved in both films where crystallites further composed of granularities; in Figure 3 coarse ballas morphology of crystallites is a combination of several spherical-shaped granularities while in Figure 3 (b) coarse ballas morphology of crystallites is a combination of several needle-shaped granularities. Due to abraded surface of seeded molybdenum –coated titanium substrate the non-uniform distribution of ballas crystallites resulted, as a result, some of the crystallites are in bigger size and some in smaller size resulting into increase the porosity of film. This change in film's morphology is mainly attributed to variation in the substrate temperature. The morphology of ballas aggregate shown in Figure 3 (a) is shown elsewhere [47] as well. The effects of substrate temperature variation under small difference were further investigated while depositing films on less abraded silicon substrates. In Figure 3 (c), the granular-like morphology is neither in tiny cluster nor in needle shape (~ 810 °C), whereas, in Figure 3 (d) the granular-like morphology altered into needle-like shape (~ 860 °C) and in Figure 3 (e) morphology transformed into lengthy needle-like shape (~ 880 °C).

In Figure 4 (a), morphology of large crystallites evolved in two different structure; tetrahedron (faces in triangle shape) and cubic (faces in square shape). In the growth of crystallites, their faces remained free from any constraint while evolving at the surface of unseeded silicon substrate where uniformly distributed pits acted as the nucleation sites. A complete detail of the process of developing pits is given elsewhere [14]. In Figure 4 (a), the growth habits of the faces of all crystallites reveal the same phenomenon owing to similar morphology where smooth faces reveal square growth while rough faces reveal triangular growth behavior; crystallites also show twin boundaries. Incorporation of impurity at any stage terminated the specific phase of evolving structure, hence, morphology at that particular node. In Figure 4 (b), crystallites developed under the same experimental conditions as in the case of Figure 4 (a) except that seeded silicon substrate involved dense growth behavior along with inter-crossing of the crystallites. Similar featured film's morphology is shown in Figure 4 (b) and also in previous studies [43, 44]. Another set of films were grown under the same strategy as in the case of Figure 4 (a) and Figure 4 (b) except that chamber pressure increased slightly and input power decreased from 5.5 to 4.7 (kW). In Figure 4 (c), the textures of crystallites don't reveal faceted faces; crystallites which developed standalone reveal different features of the faces where none of the face reveals smooth texture. Secondary nucleation is being initiated under the variation of the process parameters (gas flow rates, pressure, temperature, etc.) at fast track basis where dynamics of atoms evolving structure altered resulting into non-faceted faces of crystallites. In some crystallites, top surfaces reveal star-shaped morphology under certain mode of heat absorption (Figure 4c). Under the same parameters, film was grown on seeded silicon substrate revealing uniform and dense growth behavior along with faceted front surface (Figure 4e); the surface of crystallites evolved under atomic level controlled dynamics developing structure in square of few microns. Almost the same morphological features of crystallites are shown elsewhere [48] as in the case of Figure 4 (a) and Figure 4 (c).

In the Raman spectra of different films shown in Figure 5, the main peak at wave number 1332.1 $cm^{-1}$ is related to carbon atoms modified into diamond phase and spectra of different films show change in the peak's intensity between wave number 1350 $cm^{-1}$ to 1550 $cm^{-1}$. In Figure 5, the peak at wave number 1332 $cm^{-1}$ in each spectrum reveals carbon atoms having diamond state; in spectrum B the peak resulted in shoulder-like shape at start and end due to cube shape morphology of grains/crystallites, in spectra C and D, the peaks resulted in bell-like shape due to spherical-shaped morphology of grains/crystallites and in the spectrum E the peak resulted in Gaussian distribution shape due to Λ-shaped morphology of grains/crystallites (as in Figure 2d). Similarly, in Figure S4 different Raman spectrum shown in Figure S4 (B), Figure S4 (C), Figure S4 (D) and Figure S4 (E) are related to surface morphology of films as shown in Figure 4 (a), Figure 4 (b), Figure 4 (c) and Figure 4 (d), respectively. In Figure S4, the peak related to diamond phase has very high intensity compared to the one in Figure 5. The peak intensity is the highest in film shown in Figure 4 (b) as evolved crystallites are in pyramidal-shape morphology. In Figure 5, the large hump-like peaks at 1550 $cm^{-1}$ in spectrum C, spectrum D and spectrum E indicate modified carbon atoms in nearly graphene state as at 1580 $cm^{-1}$ graphene state exits [5]. However, peak at 1580 $cm^{-1}$ in spectrum C of Figure S4 reveals graphene state. In Figure 5 and Figure S4, the peaks more or less around 1480 cm$^{-1}$ are related to graphite structure –a master structure of carbon. Again, in all spectra shown in Figure 5 and Figure S4, minute intensity level nodes between wave number 1350 cm$^{-1}$ to 1550 cm$^{-1}$ can be observed not signifying the clear peaks are related to amorphous content of carbon. The intensive and sharp peak at 1332 cm$^{-1}$ in the Raman spectrum deduce that more carbon atoms modified into diamond state along with high purity film's region experiencing the laser beam. In the Raman spectra of different films, peaks around 1480 cm$^{-1}$ reveal carbon atoms in the graphitic structure. A small peak at 1580 cm$^{-1}$ in spectrum C of Figure S4 also reveals some carbon atoms modified into graphene state.

In synthesizing tiny grains of carbon *via* HFCVD technique the role of parameters is nearly the same as to synthesize 'tiny grains carbon thin films' *via* MPE-CVD as discussed elsewhere [5] where the main difference is the source of activation of gases (dissociation of gases). In MPE-CVD, due to incorporation of argon gas the size of tiny grains remains in the range of 1 nm to 100 nm depending on the process conditions and rate of dissociating carbon atoms from the CH$_4$. Various published high resolution transmission microscope images of 'tiny grains carbon thin films' known as UNCD/NCD films reveal that tiny grains amalgamated in the certain regions of film attained higher growth rate resulting into evolve bigger-sized grains. High resolution transmission microscope images show various features of tiny grains having atomic resolution; in some regions of the film tiny grains coalesced, in some regions they overlapped, in some regions they revealed specific orientation, in some regions disorder and in several regions tiny grains evolved two-dimensionally. Such 'tiny grains carbon thin film' deals different modification of the structure along with existence of their master structure –a graphitic structure and amorphous structure as discussed elsewhere [5].

Carbon atoms, on dissociation of methane under the thermal activation of filaments or microwave power, deposited on the substrate surface prepared under various surface treatment, in atomic from as well as in the form of tiny grains. The size of a tiny grain depends on the process parameters along with the nature of treated substrate surface. Carbon atoms amalgamated into different tiny grains under attained dynamics depending on the process conditions. Under the fixed process conditions, the modalities of binding atoms prior to deposit are the same as for those binding on deposit, however, the evolution of the structure at tiny scale may proceed in evolving different morphology-structure depending on the nature of the substrate material as well as the nature of the opted surface treatment. In the case when atoms land on the abraded surface to deposit amalgamating into tiny grains, they are in need to acquire solid state behavior to evolve either graphite structure or amorphous structure. A graphite structure at any scale is a two-dimensional structure where one set of electrons on one side of the nucleus attain neutral behavior of levitation gravitation while other set of electrons acquire neutral behavior of levitation gravitation on the other side of the nucleus, thus, binding takes place to another atom dealing similar sort of behavior under the execution of concurrent selective electron-dynamics. Shunt energy to excite an electron is the requirement in each atom to bind them under the energy knot of two photons shape-like Gaussian distribution, thus, avoiding rebound. Further detail of binding atoms of solid state execute electron transitions is discussed elsewhere [3]. However, in the case of landing, tiny grains deposit on the prepared substrate surface, atoms are not in rush to rebound as they have been bound in that size (bunch of few atoms).

When a tiny grain made under atom by atom amalgamation at substrate surface and in the region of few amalgamated atoms their executed concurrent electron-dynamics might not be uniform resulting into amorphous structure where option to modify into other phases of carbon is eliminated as the lattice become incapable to deal penetrating atomic hydrogen or atomic oxygen, however, in the structure where photon couplings between atoms remain uniform as in the case of two-dimensional structure –a graphite structure, thus, has the option to modify into either smooth elements or other phases of carbon namely, diamond, lonsdaleite and graphene while undergoing the selective gravitization of sets of electrons. The further details of their formation mechanism will be given somewhere else. The size of tiny cluster depends on the rate of binding atoms under the dissociation of methane along with localized process parameters and feed gases like molecular hydrogen and argon. In the case of HFCVD, the rate of dissociation of carbon atoms vary over the filaments, within the filaments and underneath the filaments depending on amount of employed precursor concentration along with temperature of the filaments.

The initial sticking of atoms/tiny grain at flat surface of substrate is crucial where thermal expansion coefficients do not match due to large variation of electron states of their atoms. The execution of concurrent electron-dynamics of atoms requires to be along the lateral sides and not along the normal side as carbon atoms are in graphitic state (a solid state) dealing neutral behavior of levity gravity in both sets of electrons, thus, nucleation rate of carbon atoms in graphitic state remained very slow, and particularly where substrates don't involve surface treatment or the surface treatment is poor. However, in the case of seeded/abraded surface, the initial sticking of carbon atoms/tiny grain is at fast rate starting from the nucleation as the probability of sticking atoms/tiny grain to substrate surface increased where multiple orientations protruded among arrested atoms within the cavity, crack, fracture surface or forming bunch of atoms under the time being association to atoms of seed. In such cases, the traffic of carbon atoms in graphitic state to stay in contact to surface instead of rebound largely exceeds. In Figure S5, we can observe such behaviours where crystallites grew protruding through the cavity at the surface of molybdenum-coated titanium substrate (in 'a') and growth rate was higher along the edges of unseeded silicon substrate (in 'b') revealing rich nucleation at the edges/corners of the substrate material. An ultrasonically treated surface of silicon substrate is an alternative way to achieve improved rate of nucleation where uniformly distributed pits acted as nucleation sites to deposit carbon atoms in graphitic state [14, 38]. The rate of deposition of carbon atoms in evolving certain sized grains or crystallites become uniform at the later stage of the process and their faces start formulating smooth and faceted surface as long as operating under the fixed localized process parameters. A detailed investigation of transformation of tiny grains clusters to grains of cubic morphology in multilayer strategies are discussed elsewhere [39, 49]. An abraded/seeded surface compensates the surface energy of depositing species either through trapping or rebound in cavities/pits, thus, adjusting the surface energy in terms of thermal expansion coefficient of two different materials at interface. The trafficking way of carbon atoms manipulates final structure which is undoubtedly dependent on attained dynamics at the instant of amalgamation. In some films, the entire faces of crystallites show smoothness (at atomic level) and in some not even at nanometer scale indicating varying levels of dynamics for atoms amalgamating into a grain or a crystallite. A variation in parameters alters the rate of atoms' amalgamation and influence the on-going behavior of structure evolution. To recover the phase of structure, the process may take few minutes or several minutes depending on the functionality of varied parameter (s) at that instant. Incorporation of defects at any stage of the process alters the stoichiometry of grains/crystallites of film. Regular growth combined with lateral etching has been achieved in periodic fashion where a new growth of cubic morphology grain initiated on the induction of disconnected steps in regular interval where rate of atoms' amalgamation remained at same pace by only decreasing the size of square step and coverage of atoms in each evolving grain was remained in extremely controlled manner [50].

Initially, tiny clusters formed either prior to deposit or on deposit where atoms dealt uniform electron-dynamics in evolving graphitic structure and non-uniform in evolving amorphous structure. Thus, growing tiny grain into grain, the morphology structure may alter under presently localized conditions resulting into switch graphitic phase into amorphous one or if amorphous phase was depositing switching is into graphitic phase and same is the case when a grain enlarged size in the limit of a crystallite. Now, that tiny grain, or grain or crystallite evolved into graphitic structure, in the course of evolving or after evolution might modify into fullerene structure (buckyballs, carbon nanotubes, nanostructured graphitic structure, flowery and porous structure, flake- and needle-like structure, etc.) or diamond structure, or lonsdaleite structure, or graphene structure. Originally, carbon atoms bind into two structure, namely, graphite structure (also termed as master structure or two-dimensional structure) and amorphous structure. In graphitic structure both sets of electrons reveal neutral behavior of levity gravity, one set of electrons is at one side of the nucleus while other set of electrons is at opposite side of the nucleus forming zero degree angles as positioned along the same orientation where each set of electrons constitutes three electrons as discussed elsewhere [1]. In the case of carbon atoms in gaseous states both sets of electrons are at one side of the nucleus toward North-pole [1], thus, they don't evolve any sort of structure. In the case of diamond, both sets of electrons gravitized under atomic hydrogen by forming ~ 109° between them at the centre of nucleus, in the case of lonsdaleite both sets of electrons gravitized under atomic hydrogen by forming ~ 60° angle between them at the centre of nucleus and in the case of graphene both sets of electrons gravitized under atomic oxygen by forming nearly zero degree angle between them [1]. In the case of fullerene state, one set of electrons is tacked toward North-pole and other set of electrons is tacked toward South-pole where electrons rotate and exchange positions within the margin of ~ 5° to ~ 40° angle as discussed elsewhere [1].

On increasing the substrate temperature, the morphology of tiny cluster switched into thin elliptical grain, this change in the morphology is due to attained dynamics of atoms along with heat energy. On further increasing the temperature, the rate of binding atoms in one-dimensional configuration is increased and resulting into the morphology of grain in more-like needle shape and further increase in the temperature switches morphology of tiny grain in lengthy needle-like shape. In Figure 6 ($a_1$), a layout of diamond cubic unit cell is shown where sets of electrons in each atom modified by forming ~ 109° angle with respect to centre of nucleus. In Figure 6 ($a_2$), a tiny grain mainly evolved into amorphous structure, is shown. Grain morphology in thin elliptical shape is shown in Figure 6 ($a_3$), in needle-like shape is shown in Figure 6 ($a_4$) and in lengthy needle-like shape is shown in Figure 6($a_5$). As diamond cubic unit cell has tetrahedron structure, tiny cluster has no specific structure while grain in thin elliptical shape is more on two-dimensional structure side, grain of needle-like shape switching morphology–structure from two-dimension to one-dimension while grain of long needle-like shape switched morphology–structure to one-dimension. In Figure 6 (b), two large coarse ballas morphology of crystallites are shown; in Figure 6 ($b_1$) several tiny grains morphology more-like a tiny grains are composed, whereas, in Figure 6 ($b_2$) several tiny grains morphology more-like a needle shape are composed. Such structures evolved morphology more-like in one-dimensional or two-dimensional and fall under fullerene categories of structures. In a grain or a crystallite where a particular structure switch phase from one evolving structure to another clearly reveals new morphology-structure as shown in Figure 6 (c). In Figure 6 ($d_1$), atoms with arrow lines show different attained dynamics of atoms as compared to the ones in tetrahedron unit cell. In Figure 6 ($d_2$), three tetrahedron unit cells translated tetrahedron structure on simultaneous coalescence and by preserving the symmetry as in the atoms of individual unit cell where hydrogen atoms gravitized sets of electrons of atoms, thus, formed ~ 109° angle between them at centre of nucleus. A large-sized crystallite of diamond morphology like pyramidal shape is shown Figure 6 ($d_3$). In Figure 6 ($e_1$), two-dimensional structure of grain is shown, on impinging electron streams elongated one-dimensionally where stretching of electron states remained orientational (Figure 6$e_2$). On travelling hard X-rays photons, electron states aligned resulting into modify electronic structure into smooth elements as shown in Figure 6 ($e_3$). Further details of the formation of smooth elements of tiny two-dimensional structure along with origin of $v_1$ peak have been discussed elsewhere [5].

The sticking of carbon tiny grains or binding of carbon atoms to treated or untreated substrate surface is less important in terms of thermal expansion coefficient of substrate material, which is the subject of large studies available in the literature. In fact, sticking of tiny grain or binding of atoms is more related to attained dynamics of agglomerated tiny grains or amalgamated atoms, respectively, following by electron-dynamics. As the probability of attained dynamics of carbon atoms to deposit at substrate surface is enhanced significantly when they come into solid state from gaseous where first master structure –a graphite structure evolve as long as requisite flat base for amalgamation of atoms and their binding in the scale of at least few (upto primitive cell). Neutralized behavior in terms of levity gravity of electron states of participated atoms is essential to achieve graphite structure. Certainly, the substrate material constituted of atoms of particular electron states enhance the probability of coordination to carbon atoms which are ready for depositing if retained their solid state by means of substrate atoms, they will not go anywhere. As an example, we have very less nucleation/growth of carbon films comprised grains and crystallites while dealing the flat surface but at the edges and scratched (damaged) regions of the same sample results to provide comparatively very high nucleation and growth rate as evident in Figure S5. Nevertheless, we should consider thermal expansion coefficient of suitable materials as the capability of stretching electron states of their atoms. These discussions and results enlighten us, both in terms of theory as well as in models, that there is an acute need to revisit nucleation and growth mechanisms in films of different sorts of materials at impartial grounds.

The process of synergy also play vital role to design material of tiny grains, grains, and crystallites in the form of films in various deposition technique as it has strong impact on both the attained dynamics of tiny grains/atoms as well as electron-dynamics and then travelling photons available in the medium under varying wavelength, energy and shape as discussed elsewhere in the case of tiny metallic colloids [2]. There is a need to automatize such kind of parameters as they directly and indirectly influence the certain growing phase of structure, thereby, vary the performance of carbon-based materials while using them for various characterization and analysis tools and applications.

## 4. CONCLUSIONS

In carbon materials, both atoms' amalgamation and tiny grains' amalgamation under attained dynamics resulting in evolving either graphite structure or amorphous structure where binding is under uniform electron-dynamics or non-uniform electron-dynamics, respectively. Carbon atoms, when in solid state, amalgamate into tiny grains, grains and crystallites as per attained dynamics and electron-dynamics are the origin of their morphology-structure. In both materials processing techniques, binding of atoms into tiny grain and formation of diverse morphology structure of films is related to localized process conditions. Within the same process, earlier nucleated grains possess different morphology than those which grow later. The morphology of evolving structure alters within short-range, moderate-range or long-range orders depending on the attained dynamics of amalgamating atoms. Because different growth behavior of carbon materials achieved under several conditions of parameters in two different techniques, it is possible that growth behavior of certain phase is controlled by different physical behavior under large variation of set parameters. Then, a carbon atom, when in solid state, has the option to switch into various state (phase) under the behavior of field force, thus, influencing the morphology-structure under ordinary taken measures. Under the slight variation of the localized parameter (s), a tiny grain influences the structure by switching it into another one. An abraded substrate surface provides the maximum probability of binding atoms under attained dynamics resulting into fast rate nucleation of depositing carbon atoms, hence, improve growth rate of grains and crystallites of film. The Raman peaks trends in deposited films under varying process parameters validate the presence of different morphology-structure (phases) of carbon. The present study sets new trends in synthesizing carbon materials and solicits reinvestigation of their performance, at nanoscale and microns, by underpinning explanations in relation to evolved structure along with modified structure under the behavior of field force, thus, morphology-structure under varying conditions of the process.

**Acknowledgements:**

Mubarak Ali expresses his sincere gratitude to The Scientific and Technological Research Council of Turkey (TÜBİTAK, letter ref. # B.02.1.TBT.0.06.01-216.01-677-6045) for honoring postdoctorship (year 2010). Mubarak Ali thanks Professor I-Nan Lin, Tamkang University, Taiwan (R.O.C.) for useful discussions on 'tiny grain carbon thin films' while working as postodoc candidate (August 2013-July 2014). Mubarak Ali heartily thanks Mr. Talat ALPAK (ITU, Istanbul) in field emission scanning microscope operation and appreciates useful discussion with Dr. M. Ashraf Atta.

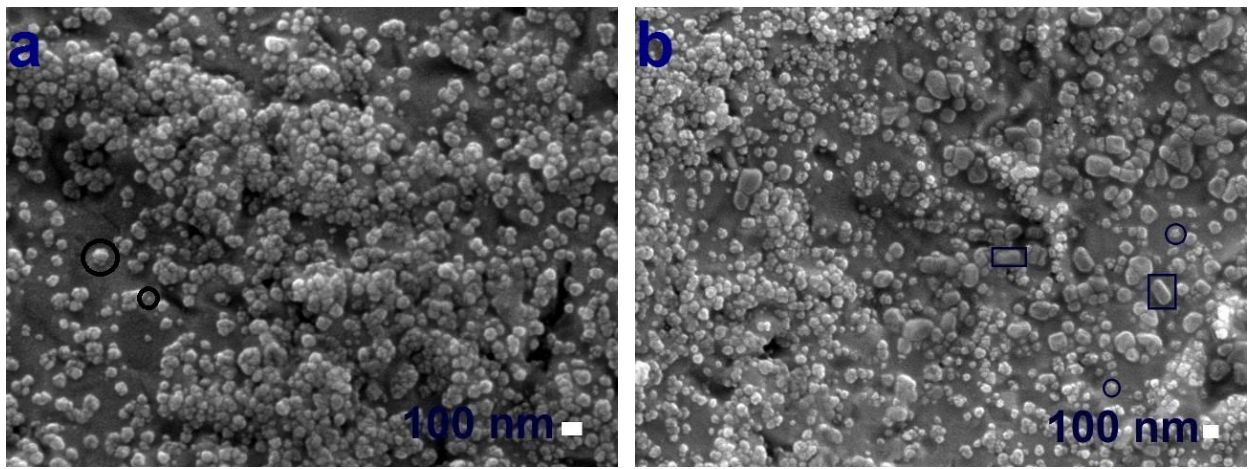

**Figure 1:** Scanning microscope surface images of tiny clusters deposited on seeded Mo-coated Ti substrate (SAMPLE A)

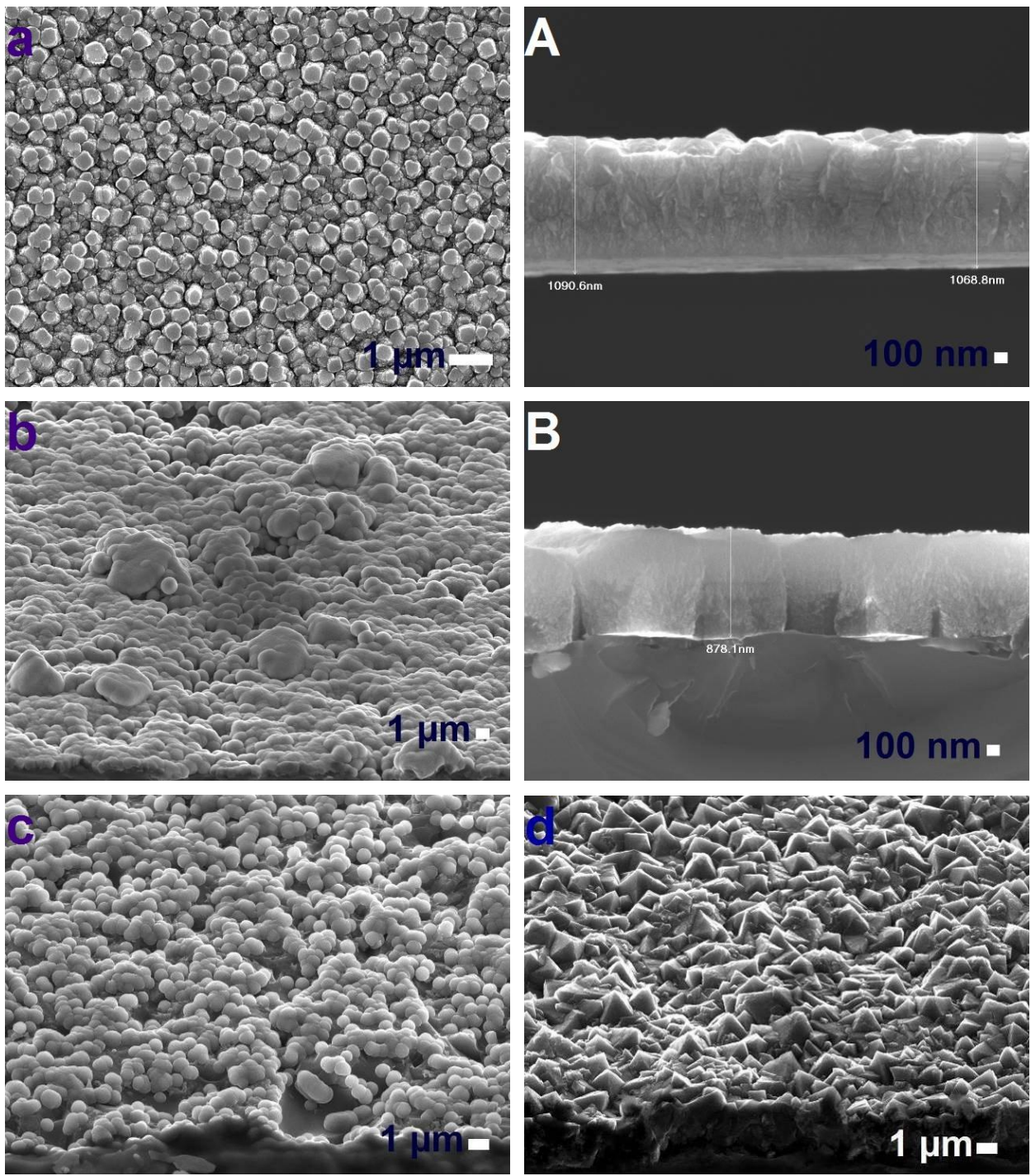

**Figure 2:** **(a)** Scanning microscope surface and fracture cross-section images of film deposited on seeded Si substrate (SAMPLE B), **(b)** scanning microscope surface and fracture cross-section images of film deposited on seeded Si substrate (SAMPLE C), **(c)** scanning microscope image taken from the tilted position shows both surface and cross-section of deposited film on seeded Si substrate (SAMPLE D) and **(d)** scanning microscope image taken from the tilted position shows both surface and cross-section of deposited film on seeded Si substrate (SAMPLE E).

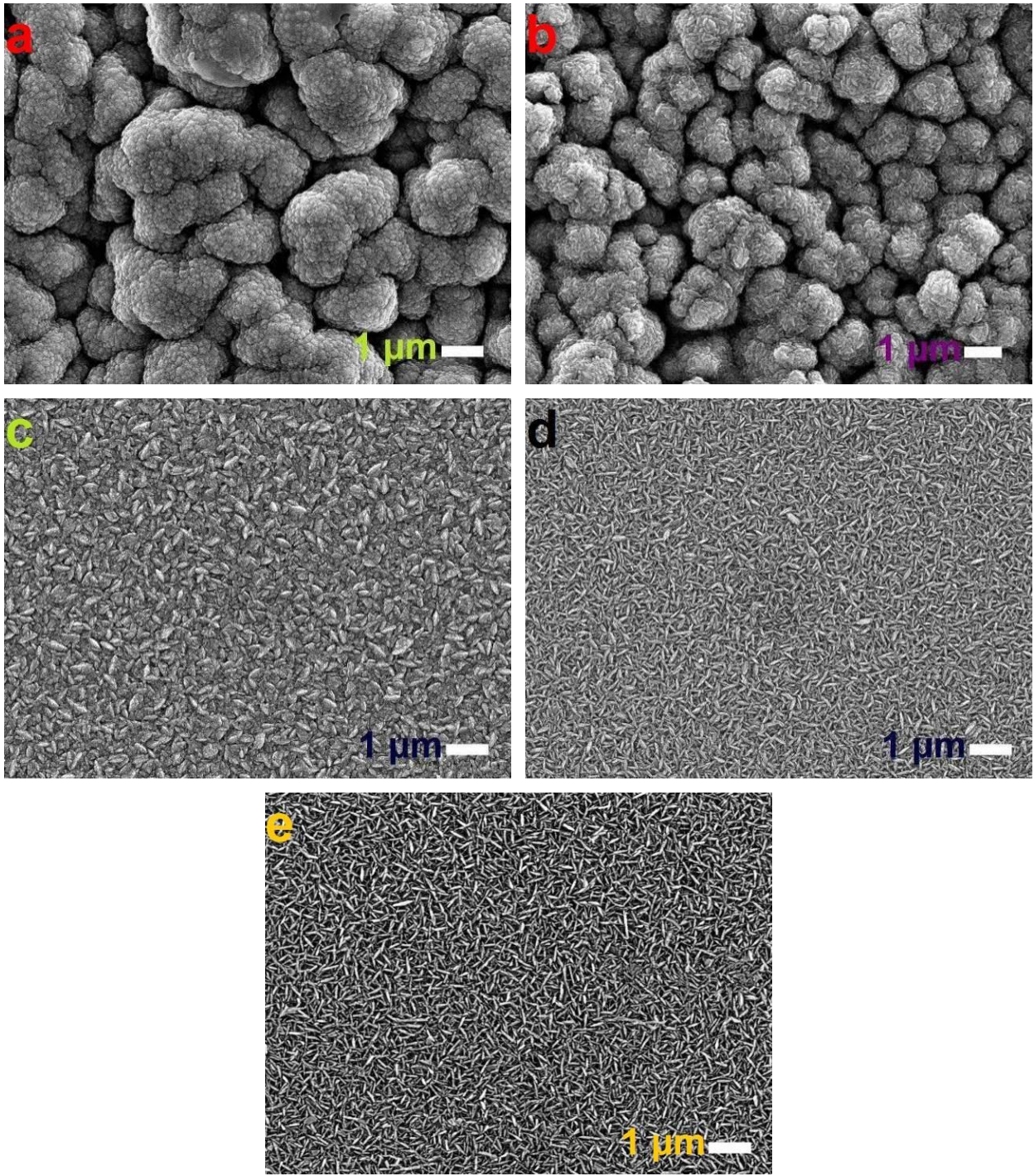

**Figure 3:** Scanning microscope surface image of large ballas crystallites deposited on seeded Mo-coated Ti substrate morphology of grains more like **(a)** tiny grain and **(b)** tiny needles, scanning microscope surface image of film deposited on seeded Si substrate shows morphology like **(c)** uniformly distributed thin elliptical grains, **(d)** uniformly distributed needle-like grains and **(e)** uniformly distributed lengthy needle-like grains (SAMPLE F).

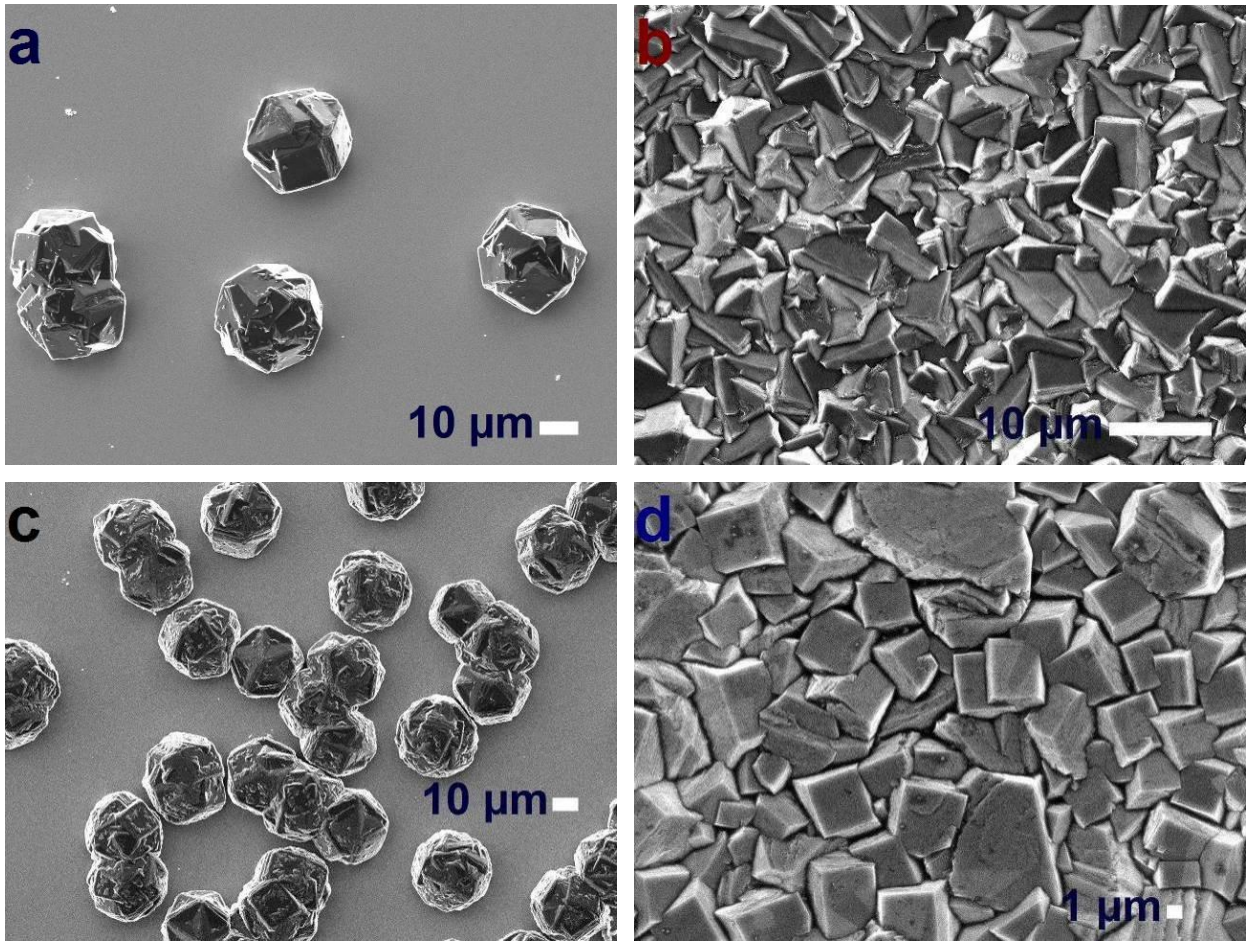

**Figure 4:** Scanning microscope surface image of large crystallites deposited on (a) unseeded Si substrate and (b) seeded Si substrate (SAMPLE G).

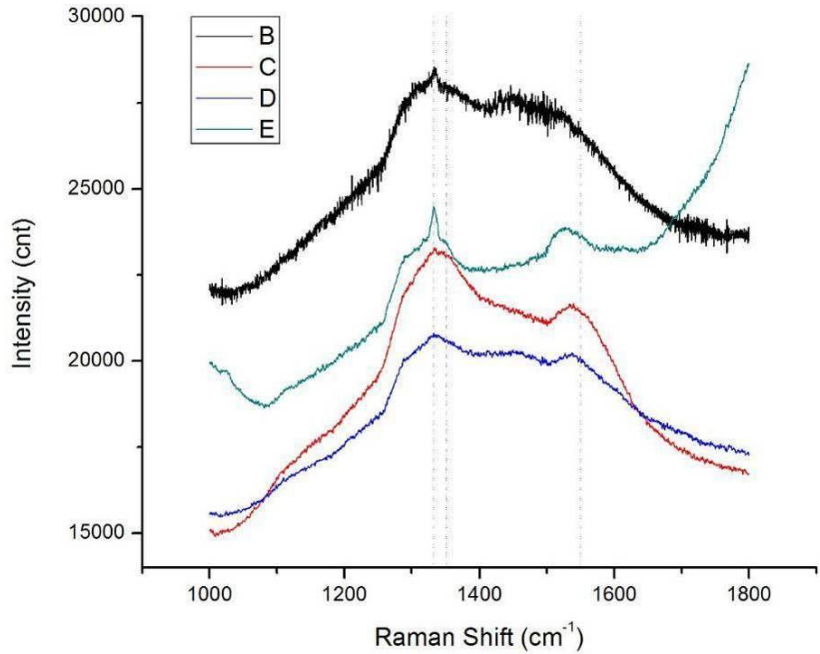

**Figure 5:** Raman spectrum (B), (C), (D) & (E) of films in Figure 2 (a), Figure 2 (b), Figure 2 (c) and Figure 2 (d).

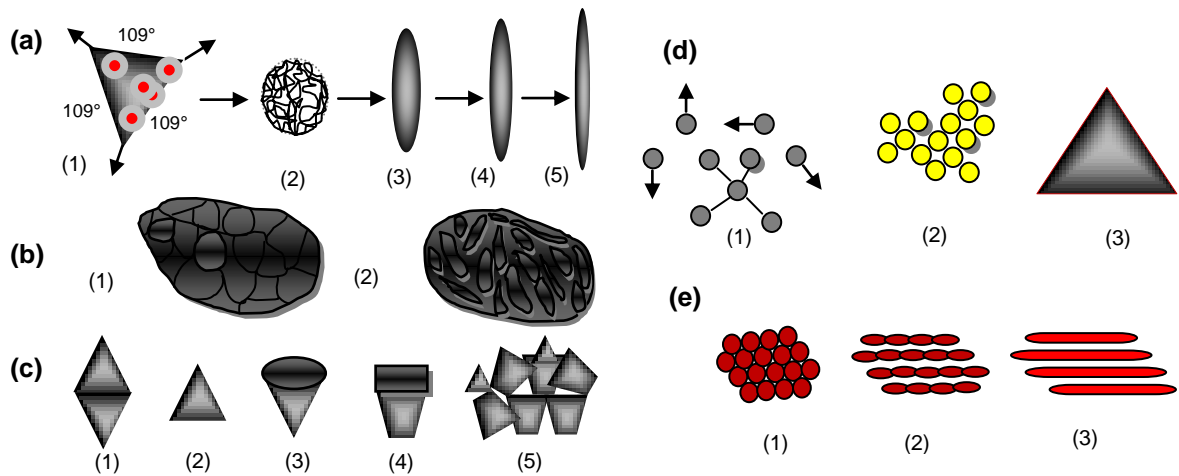

**Figure 6:** ($a_1$) diamond primitive cell, ($a_2$) amalgamation of tiny grains and their sticking, ($a_3$) grain in thin elliptical morphology, ($a_4$) grain in needle-like shape and ($a_5$) grain in lengthy needle-like shape, ($b_1$) amalgamation of several tiny grains (morphology like tiny cluster) into coarse ballas morphology of crystallite and ($b_2$) amalgamation of several tiny grains (morphology like needle) into coarse ballas morphology of crystallite, ($c_1$) bottom-to-bottom binding of tetrahedron structure in pyramidal-like morphology, ($c_2$) tetrahedron structure of grain/crystallite, ($c_3$) transformation of tetrahedron structure into two-dimensional structure, ($c_4$) transformation of tetrahedron structure in cube shape (cubic structure) and ($c_5$) contaminated structure of crystallite comprising different phases, ($d_1$) modification of amalgamated carbon atoms into tetrahedron primitive cell, ($d_2$) binding of tetrahedron primitive cells under identical attained dynamics, ($d_3$) crystallite of tetrahedron structure revealing pyramidal-like morphology, ($e_1$) two-dimensional structure, ($e_2$) elongation of two-dimensional structure and ($e_3$) formation of smooth elements while travelling photons of hard X-rays and their binding under low energy photons.

## Authors' biography:

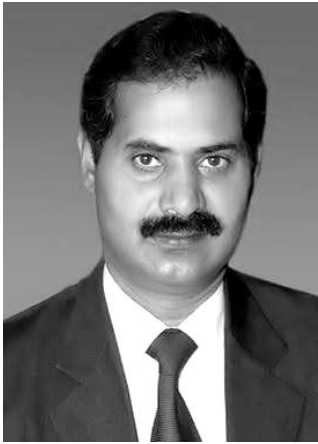

Mubarak Ali graduated from University of the Punjab with B.Sc. (Phys& Maths) in 1996 and M.Sc. Materials Science with distinction at Bahauddin Zakariya University, Multan, Pakistan (1998); thesis work completed at Quaid-i-Azam University Islamabad. He gained Ph.D. in Mechanical Engineering from Universiti Teknologi Malaysia under the award of Malaysian Technical Cooperation Programme (MTCP;2004-07) and postdoc in advanced surface technologies at Istanbul Technical University under the foreign fellowship of The Scientific and Technological Research Council of Turkey (TÜBİTAK; 2010). He completed another postdoc in the field of nanotechnology at Tamkang University Taipei (2013-2014) sponsored by National Science Council now M/o Science and Technology, Taiwan (R.O.C.). Presently, he is working as Assistant Professor on tenure track at COMSATS Institute of Information Technology, Islamabad campus, Pakistan (since May 2008) and prior to that worked as assistant director/deputy director at M/o Science & Technology (Pakistan Council of Renewable Energy Technologies, Islamabad; 2000-2008). He was invited by Institute for Materials Research (IMR), Tohoku University, Japan to deliver scientific talk on growth of synthetic diamond without seeding treatment and synthesis of tantalum carbide. He gave several scientific talks in various countries. His core area of research includes materials science, physics & nanotechnology. He was also offered the merit scholarship (for PhD study) by the Government of Pakistan but he couldn't avail. He is author of several articles published in various periodicals (https://scholar.google.com.pk/citations?hl=en&user=UYjvhDwAAAAJ).

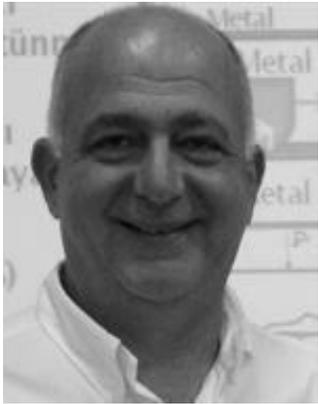

Mustafa Ürgen graduated from Istanbul Technical University in 1977, Mining Faculty, followed by his M.Sc. from Metallurgical Engineering Faculty in 1978 and PhD from Institute of Science and Technology in 1986. He remained visiting fellow at Max-Planck Institut, Institut für Metallwissenschaften, Stutgart, Germany (1988-89) and won IMF award – Jim Kape Memrial Medal in1998. He is one of the top grant winning senior standing professors of the Istanbul Technical University. Professor Ürgen is head of several laboratories in diversified areas of science & technology (Surface Technologies: Electrolytic and conversion coatings, diffusion coating techniques, vacuum coating techniques, PVD coatings, composite coatings, surface analysis. Corrosion and Corrosion Protection: Mechanism of corrosion reactions, pitting corrosion (stainless steel, aluminum alloys, ceramic, DLC coated metals, stress corrosion cracking). He worked at several managerial and administrative positions at departmental, faculty and university levels and supervised several PhD and postdoc candidates funded locally as well as internationally, and some of his students are working as a full professor. Professor Ürgen delivered many talks at various forums, to his credit. He secured commercialized patents and long list of publications in referred journals diversify class of materials, physics and chemistry and other interdisciplinary areas of science and technology.

## Supplementary Materials:

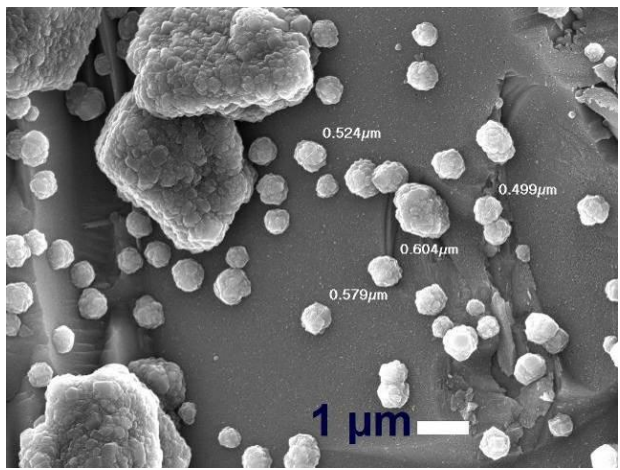

**Figure S1:** Scanning microscope surface image of cauliflower-like morphology of micron/submicron-sized isolated clusters deposited on seeded Si substrate; total mass flow rate: 200 sccm (0.75% $CH_4$), chamber pressure: ~ 55 torr, growth time: 8 hours (nucleation time: 30 minutes where concentration of $CH_4$ was kept 1% and chamber pressure increased from 10 torr to 37 torr), distance b/w substrate and hot filaments: ~ 6.0 mm, substrate temperature: 700 ± 20°C, input power: ~ 2.7 (kW) and ultrasonic agitation of substrate for 5 minutes in acetone having 28 μm sized diamond powder (70 ml:7 gram) following by mechanically scratching to 5 μm sized diamond powder slurry and 10 minutes wash with acetone in ultrasonic bath.

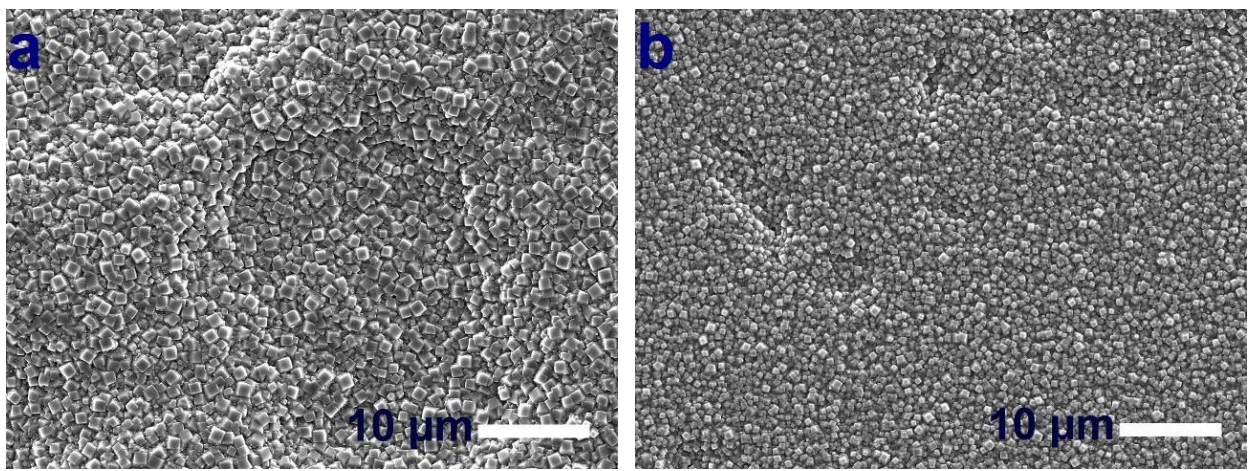

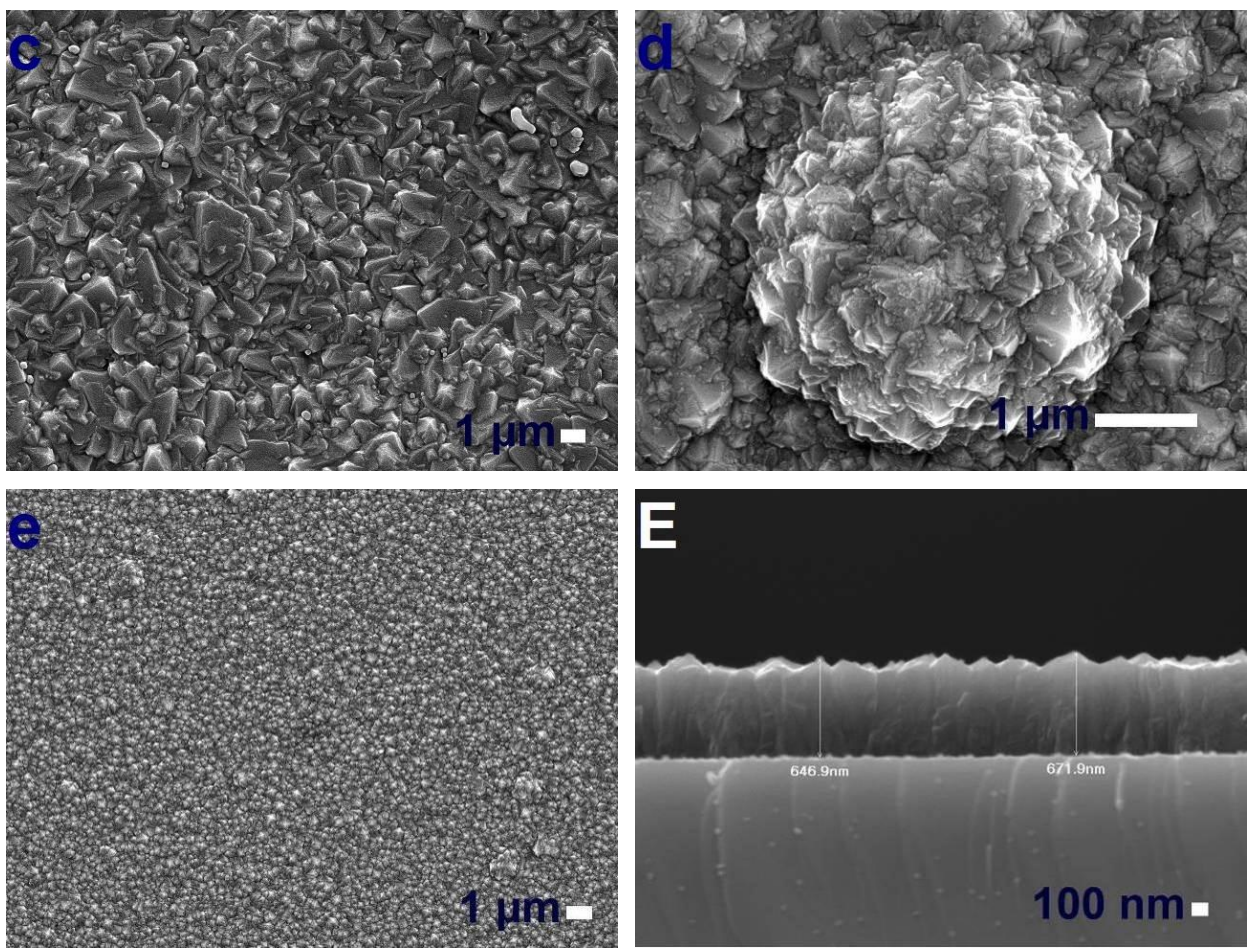

**Figure S2:** Scanning microscope surface image of film having cubic morphology of crystallites deposited **(a)** on seeded Mo-coated Ti substrate and **(b)** on seeded Si substrate; total mass flow rate: 200 sccm (0.75% $CH_4$), chamber pressure: ~ 40 torr, growth time: 9 hours 30 minutes (nucleation time: 30 minutes where concentration of $CH_4$ was kept 1% and pressure increased from 10 torr to 40 torr), distance b/w substrates and hot filaments: ~ 7.0 mm, substrate temperature: 830 ± 20°C, input power: ~ 3 (kW) and for 90 minutes ultrasonic agitation of substrate with 500-600 mesh diamond powder in acetone (7 gram:70 ml) following by 10 minutes washing in acetone; **(c)** Scanning microscope surface image of pyramidal-shaped/broken faces crystallites of film deposited on seeded Mo-coated Ti substrate, **(d)** Scanning microscope surface image of large protruded pyramidal-shaped crystallite deposited on seeded Mo-coated Ti substrate, **(e)** Scanning microscope surface image of uniformly distributed pyramidal-shaped crystallites deposited on seeded Si substrate and **(E)** Scanning microscope fracture cross-section image of film (in columnar growth) deposited on seeded Si substrate; total mass flow rate: 200 sccm (1.25% $CH_4$), chamber pressure: ~ 30 torr, growth time: 6 hours 30 minutes (nucleation time: 30 minutes where concentration of $CH_4$ was kept 2% and pressure increased from 10 torr to 30 torr), distance b/w substrates and hot filaments: ~ 7.0 mm, substrate temperature: 780 ± 20°C, input power: ~ 2.7 (kW) and ultrasonic agitation of substrate for 65 minutes with 500-600 mesh diamond powder in acetone (8 gram:80 ml) and 10 minutes wash with acetone in ultrasonic bath.

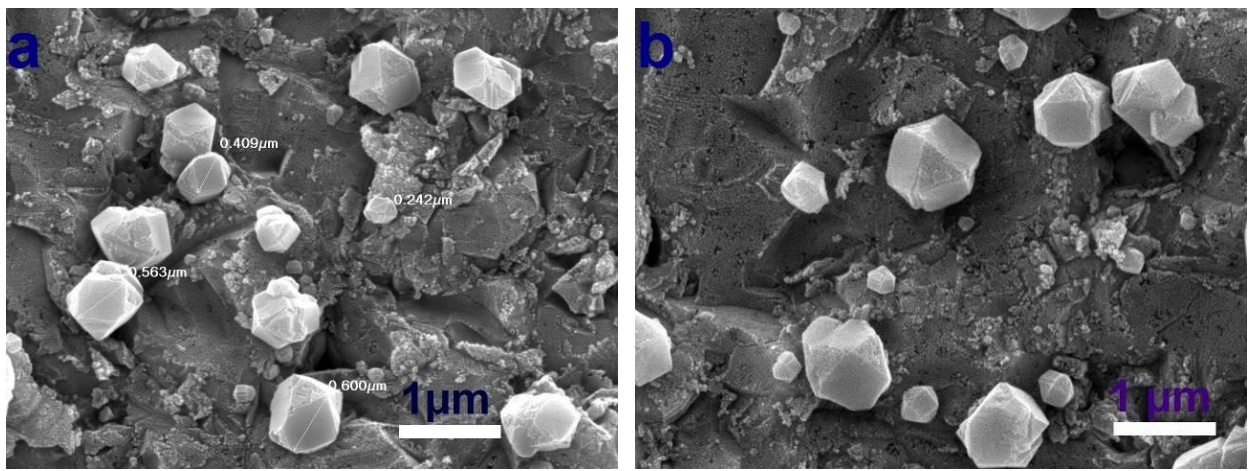

**Figure S3:** (a & b) Scanning microscope surface images of grains/crystallites grown on abraded silicon substrate under seeding treatment; total mass flow rate: 300 sccm (0.75% $CH_4$), chamber pressure: ~ 40 torr, growth time: 9 hours 30 minutes (nucleation time: 30 minutes where concentration of $CH_4$ was kept 1% and pressure increased from 10 torr to 40 torr), distance b/w substrates and hot filaments: ~ 8.0 mm, substrate temperature: 900 ± 20°C, input power: ~ 3.7 (kW) and 5 minutes mechanically scratch of substrate with 28 microns diamond powder suspension (in acetone) following by mechanically scratching for 10 minutes with 5 microns diamond powder suspension (in acetone) and 10 minutes wash with acetone in ultrasonic bath, again the substrate ultrasonically agitated for 90 minutes with 500-600 mesh diamond powder in acetone solution (7 gram:70 ml) and 10 minutes wash with acetone in ultrasonic bath.

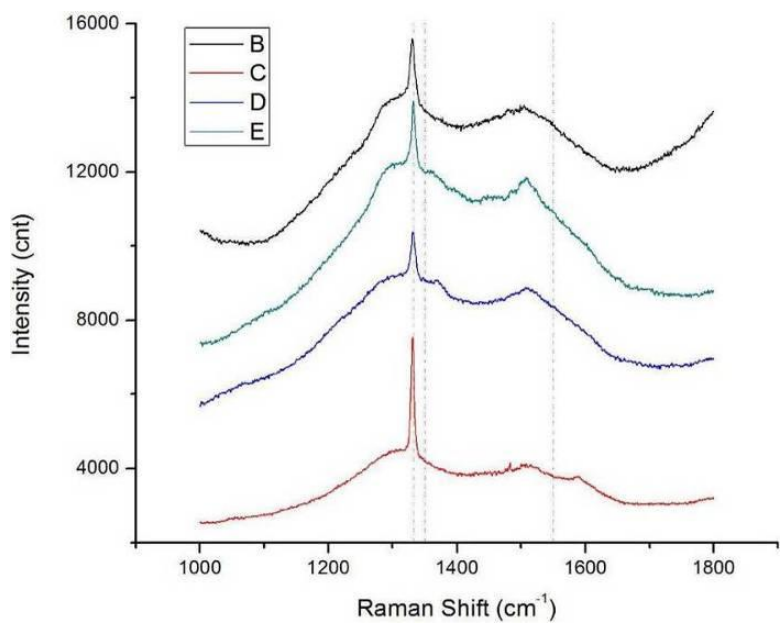

**Figure S4:** Raman spectrum (B), (C), (D) & (E) of films for which surface morphology is shown in Figure 4 (a), Figure 4 (b), Figure 4 (c) and Figure 4 (d), respectively.

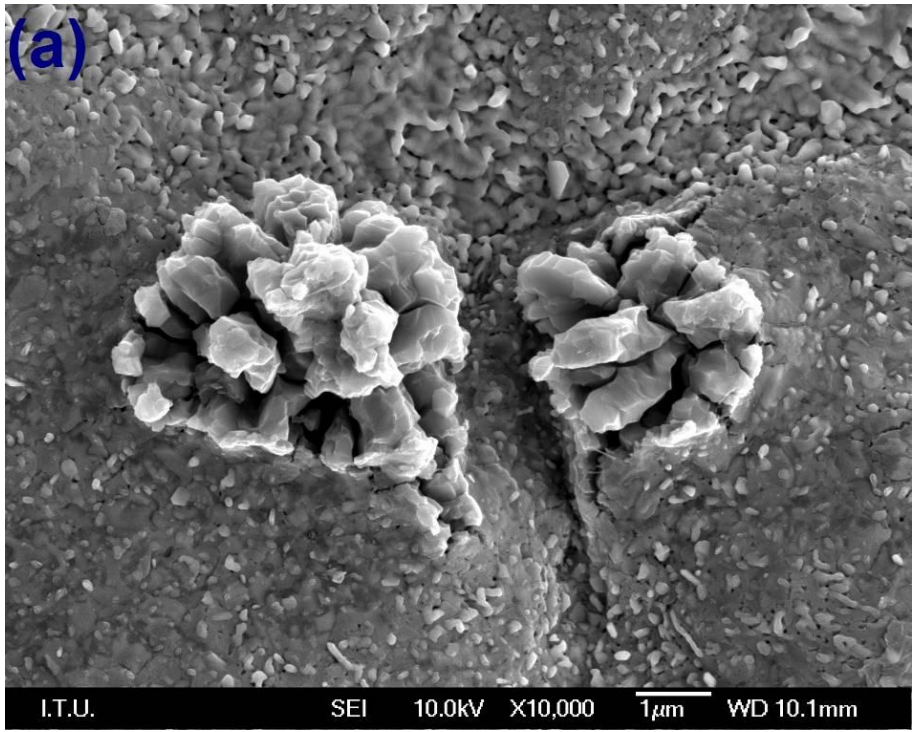
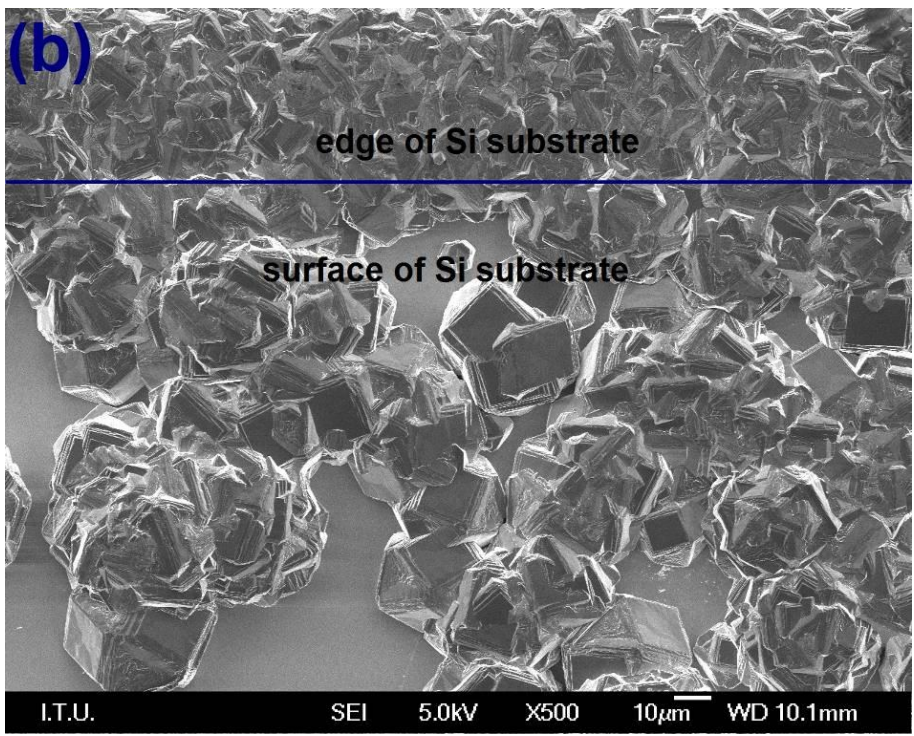

**Figure S5:** Scanning microscope surface images of (a) diamond crystallites protruding through the cavity at Mo-coated Ti substrate while evolution and (b) diamond grains and crystallites evolved on unseeded Si substrate where growth rate is very high along the edge.